\documentclass[final,5p,times,twocolumn]{elsarticle}

\usepackage[T1]{fontenc}
\usepackage[utf8]{inputenc}

\usepackage{amsmath}
\usepackage{amsfonts}
\usepackage{amssymb}
\usepackage{amsxtra}
\usepackage{array}
\usepackage{graphicx}
\usepackage{hepunits}
\usepackage{units}
\usepackage{color}
\usepackage{hyperref}
\usepackage{xspace}
\usepackage[dvipsnames]{xcolor}
\usepackage[normalem]{ulem}

\usepackage[mathscr,scaled=1.15]{urwchancal}

\DeclareMathOperator\arctanh{arctanh}

\newcommand{\refeq}[1]{Eq.~\eqref{#1}}
\newcommand{\refeqs}[2]{\textrm{Eqs}.~(\ref{#1})-(\ref{#2})}

\newcommand{\reffig}[1]{Fig.~\ref{#1}}
\newcommand{\refcite}[1]{Ref.~\cite{#1}}
\newcommand{\refscite}[1]{Refs.~\cite{#1}}
\newcommand{\refsec}[1]{Sec.~\ref{#1}}

\newcommand{\eg}{\textit{e.g. }}

\newcommand{\vperp}[1]{\ensuremath{\textbf{#1}_\perp}}

\newcommand{\HPupi}[0]{{H_{\mathrm{P}}}_{\pi^+}^u}

\biboptions{sort&compress}

\journal{Physics Letters B}

\begin{document}

\begin{frontmatter}

\title{A Nakanishi-based model illustrating the covariant extension of the pion GPD overlap representation and its ambiguities}

\author[CEA]{N. Chouika}
\author[INFN]{C. Mezrag} 
\author[CEA]{H. Moutarde}
\author[Huelva,CAFPE]{J.~Rodr\'iguez-Quintero}
\address[CEA]{IRFU, CEA, Université Paris-Saclay, F-91191 Gif-sur-Yvette, France}
\address[INFN]{Istituto Nazionale di Fisica Nucleare, Sezione di Roma, P. le A. Moro 2, I-00185 Roma, Italy}
\address[Huelva]{Dpto. Ciencias Integradas, Centro de Estudios Avanzados en Fis., Mat. y Comp., Fac. Ciencias Experimentales, Universidad de Huelva, Huelva 21071, Spain}
\address[CAFPE]{CAFPE, Universidad de Granada, E-18071 Granada, Spain}

\begin{abstract}
A systematic approach for the model building of Generalized Parton Distributions (GPDs), based on their overlap representation within the DGLAP kinematic region and a further covariant extension to the ERBL one, is applied to the valence-quark pion's case, using light-front wave functions inspired by the Nakanishi representation of the pion Bethe-Salpeter amplitudes (BSA).
This simple but fruitful pion GPD model illustrates the general model building technique and, in addition, allows for the ambiguities related to the covariant extension, grounded on the Double Distribution (DD) representation, to be constrained by requiring a soft-pion theorem to be properly observed. 
\end{abstract}


\end{frontmatter}

\section{Introduction}

GPDs provide a three-dimensional picture of hadrons~\cite{Burkardt:2000za}, unifying both Parton Distributions Functions (PDFs) and Form Factors into a single nonperturbative object which yields information about the distributions of partons within the light front.
 After their introduction 20 years ago~\cite{Mueller:1998fv, Ji:1996nm, Radyushkin:1997ki}, GPDs became a hot topic in hadron physics which many experimental and theoretical efforts have been since then devoted to (see \eg~\refscite{Ji:1998pc,Goeke:2001tz, Diehl:2003ny, Belitsky:2005qn, Boffi:2007yc, Guidal:2013rya,Mueller:2014hsa}).
Still today, they constitute a central goal contributing to guide experimental programs, within the framework of an international cooperative effort addressed to the understanding of the deep internal structure of hadrons on the basis of QCD. In order to gain insight into this internal structure, the appropriate description of GPDs plays an essential role. 

To this purpose, either following a purely phenomenological approach~\cite{Guidal:2004nd,Goloskokov:2005sd, Polyakov:2008aa, Kumericki:2009uq, Goldstein:2010gu, Mezrag:2013mya} or handling a nonperturbative framework that might possess a direct connection with QCD (see \eg~\refscite{Mezrag:2016hnp, Fanelli:2016aqc, Dorokhov:2011ew,Broniowski:2007si,Tiburzi:2017brq} and references therein), some genuine constraints should be crucially observed.
In particular, any theoretical construction properly endowed for an accurate extrapolation of the experimental GPD information is challenged by the need to fulfill the {\it polynomiality} and {\it positivity} properties. 
{\it Positivity} is a quantum mechanics implication which results from the positivity of the norm in a Hilbert space, while {\it polynomiality} is the consequence of the Lorentz invariance in a quantum field theory, both very fundamental properties grounded on the underlying structure and symmetries of QCD.
 Only in very few cases, as \eg~\refcite{Broniowski:2007si}, particular models have been developed by taking care of both properties simultaneously.
 More often, building a GPD model or applying a given computational technique implies to favor one or the other, with no guarantee for both being respected at the same footing.
 Nevertheless, an interesting approach was pioneered by the authors of \refcite{Hwang:2007tb}, based on the GPD overlap representation, guaranteeing positivity, and its further covariant extension, respecting polynomiality, guided by the Double Distribution representation.
 However, the technique was developed only for a specific algebraic model of light-front wave functions (LFWFs). 
 We generalized it recently in a model-independent way based on the Radon inverse transform in~\refcite{Chouika:2017dhe} and lengthily discussed therein a fully systematic technique to achieve that goal.
 It is worth noting that another technique based on the inverse Laplace transform has been more recently presented in \refcite{Muller:2017wms}.
 The basic ingredient for implementing our method is the knowledge of the LFWF for the hadron, whichever model or computational framework might be employed to obtain it.
 This letter is particularly intended to illustrate this technique with its application to the LFWFs derived from a pion Bethe-Salpeter amplitude (BSA) based on the Nakanishi representation~\cite{Nakanishi:1969ph,Nakanishi:1963zz} in~\refscite{Mezrag:2014jka,Mezrag:2016hnp} and to the pion DGLAP GPD therein developed.
 But, specially, we also deal here with the ambiguities related to the covariant extension to the ERBL region, by using a soft-pion theorem~\cite{Polyakov:1998ze} for their constraining, and thus produce a full sketch of the pion valence-quark GPD based on the Bethe-Salpeter LFWFs.

\section{The covariant extension of the GPD overlap representation: generalities \label{sec:gpd-theory-modeling}}

Let us here briefly sketch the approach of \refcite{Chouika:2017dhe} for the covariant extension of GPDs obtained in the overlap representation from DGLAP to ERBL kinematical domains, specially emphasizing the resulting ambiguities.

GPDs are defined as a lightfront projection of a non-diagonal hadronic matrix element of a bi-local operator.
For instance, the twist-2 chiral-even quark GPD of a pion can be written as follows:
\begin{eqnarray}
  \label{eq:GPDDefinition}
  && H^{q}\left(x,\xi,t\right) = \frac{1}{2}\int\frac{\mathrm{d}z^{-}}{2\,\pi}\,e^{i\,x\,P^{+}z^{-}} \\ 
 && \times \left.\left\langle
  P+\frac{\Delta}{2}\right|\bar{\psi}^q\left(- \frac z 2\right)\gamma^{+}\psi^q\left(\frac z 2\right)\left|P-\frac{\Delta}{2}\right\rangle
  \right|_{z^{+}=0,\,z_{\perp}=0}\,, \nonumber 
\end{eqnarray}
where $P$ (resp. $\Delta$) is the momentum average (resp. transfer) of the hadron states, $t=\Delta^{2}$ and $x$ (resp. $\xi=-\frac{\Delta^{+}}{2\,P^{+}}$) is the longitudinal momentum fraction average (resp. transfer) of the quarks ($q$ classically stands for the quark flavor). 
Due to time reversal invariance, the so defined GPDs are even in $\xi$ and we will then restrict to $\xi \geq 0$ in the following (unless explicitly stated otherwise). 
PDFs can be recovered from GPDs as their forward limit, $\Delta = 0$, while the hadron elastic form factor can be expressed as a GPD sum rule. 
A bridge between PDFs and hadron form factors is thus paved by GPDs.
We will further insist on this as a first benchmark for the construction of the GPD model.  

On the other hand, it is well known that lightfront quantization allows the expansion of any hadron state of given momentum and polarization on a Fock basis of  N-particles partonic states, weighted by the so-called lightfront wave functions (LFWFs) which contain all the nonperturbative physics \cite{Brodsky:1997de}. 
Thus, one can express GPDs in terms of LFWFs \cite{Diehl:2000xz}, albeit the partonic picture and therefore the way the GPDs and LFWFs relate to each other depend on the considered kinematics. 

In the so-called DGLAP region ($|x|\ge \xi$), the GPD is given by an overlap of LFWFs defined for the \emph{same} number of constituents. 
In particular, keeping the example of the pion and restraining ourselves to the valence contribution ({\it i.e.} the two-particle Fock sector), in the region $x \ge \xi$, we have \cite{Diehl:2003ny}:
\begin{eqnarray}
H_{\pi^{+}}^{u}\left(x,\xi,t\right)=\int\frac{\mathrm{d}^{2}\mathbf{k}_{\bot}}{16\,\pi^{3}}
\Psi_{u\bar{d}}^{*}\left(\frac{x-\xi}{1-\xi},\mathbf{k}_{\bot}+\frac{1-x}{1-\xi} \frac{\Delta_\bot}{2}\right) \nonumber \\ 
\times \; \Psi_{u\bar{d}}\left(\frac{x+\xi}{1+\xi},\mathbf{k}_{\bot}-\frac{1-x}{1+\xi} \frac{\Delta_\bot}{2}\right)\,,\label{eq:overlap_ud}
\end{eqnarray}
where, specializing to the $\pi^{+}$ case, $\Psi$ is the pion LFWF for the $u\bar{d}$ two-particle Fock sector. 
\refeq{eq:overlap_ud} provides us with a two-particle truncated expression for the pion GPD in the DGLAP kinematic domain, which highlights the underlying Hilbert space structure and makes possible to show the above-mentioned positivity property \cite{Pire:1998nw,Diehl:2000xz, Pobylitsa:2001nt, Pobylitsa:2002gw}. 

The GPD can be also generally derived in the other kinematic domain, called ERBL ($\xi \ge |x|$), following the same overlap approach but then involving LFWFs for \emph{different} numbers of constituents, namely $N$ and $N+2$.
 Thereupon, in our pion special case, no two-particle truncated expression suits within the overlap representation, as the first non-vanishing contribution to the GPD will result from the overlap of LFWFs defined for the 2- and 4-particle Fock sectors.
 Indeed, the latter reflects a more general and deeper feature: inasmuch as independent descriptions of the DGLAP and ERBL regions will almost certainly break polynomiality (as stressed, for instance, in \refcite{Diehl:2003ny}), the observance of the Lorentz covariance will result from a delicate compensation of contributions to the GPD's Mellin moments from both DGLAP and ERBL regions.
 Therefore, Lorentz invariance strongly ties $N$- to  $(N+2)$-particle LFWFs, in general, and 2- to 4-particle ones, in our special case, thus preventing from a consistently covariant description, in the overlap representation, for the valence-quark GPD approximated within the lowest Fock-basis sector. 

In particular cases, covariant extensions of an overlap of LFWFs from the DGLAP to the ERBL region can be found in the literature~\cite{Hwang:2007tb, Mueller:2014tqa}. 
As mentioned above, we have recently presented~\cite{Chouika:2017dhe} a general solution to this problem on the mathematical ground of a natural expression for the polynomiality condition: the Double Distribution (DD) representation of the GPD.
 The polynomiality property is expressed by the condition that the GPD's $m{\text{th}}$-order Mellin moment is a $(m+1)$-degree polynomial in the skewness variable, $\int_{-1}^{1}dx x^m H(x,\xi,t) = \sum_{k=0}^{m+1}c^{(m)}_k(t) \xi^k$, for all non-negative integers $m$.
 Let us now assume that there exists\footnote{The function $D$ appears thus defined, at any value of $t$, by all their Mellin moments.} a function $D(x,t)$ with support $x \in [-1,1]$ for any $t$ such that $\int_{-1}^{1}dx x^m D(x,t)=c^{(m)}_{m+1}(t)$, in such a way that the $m{\text{th}}$-order Mellin moments of $H(x,\xi,t)-\mathrm{sgn}(\xi) D(x/\xi,t)$ are polynomials of degree $m$ in $\xi$.
 It can be thereupon formally and rigorously concluded~\cite{Chouika:2017dhe,Hertle:1983} that $H(x,\xi,t)-\mathrm{sgn}(\xi) D(x/\xi,t)$ results from the Radon transform~\cite{Teryaev:2001qm,Radon:1917tr} of a given distribution $F_D$\footnote{The $\mathrm{sgn}(\xi)$ comes from the jacobian of the change of variables $x \to x/\xi$ and makes transparent the $\alpha$-parity of $D(\alpha,t)$ when $\xi < 0$. The same happens for $D^+$ and $D^-$ in \eqref{eq:gpd-P-Ds}. This is why we consider there and here the general case $\xi \in [-1,1]$.}, 
%---
\begin{equation}
\label{eq:radon-transform-in-dd-plus-d-gauge}
H(x, \xi,t) - \mathrm{sgn}(\xi) D(x/\xi,t) = \int_\Omega \mathrm{d}\beta \mathrm{d}\alpha \,
F_D(\beta, \alpha,t) \delta(x - \beta - \alpha \xi) \; ,
\end{equation}
where the support $\Omega = \left\{\left(\beta,\alpha\right) \in \mathbb{R}^2 / \,
|\beta|+|\alpha| \le 1\right\}$ reflects the physical domain of GPDs $(x, \xi \in [-1, +1]$).
 It should be noticed that, $H(x,\xi,t)$ being even in $\xi$, $F_D(\beta,\alpha,t)$ is an even function in $\alpha$, $D(x,t)$ is odd in $x$ and, accordingly, $c^{(m)}_{m+1}(t)\equiv 0$ for any even integer $m$.
 In particular, $c^{(0)}_1(t)\equiv 0$ and thus $\int_{-1}^1 H(x,\xi,t) = c^{(0)}_0(t)$, not depending on $\xi$ as the form factor sum rule requires. 
Furthermore, $D(x/\xi,t)/|\xi|$ is the Radon transform of the distribution $G_D(\beta,\alpha,t)=D(\alpha,t) \delta(\beta)$ and, on top of this, according to \cite{Teryaev:2001qm,Tiburzi:2004qr}, the pair of distributions $(F_D,G_D)$ does not constitute a unique parametrization for the integral representation of the same given GPD but it can be transformed in a new couple $(F,G)$ such that
%-----
\begin{equation}
\label{eq:gpd-in-general-F-and-G-gauge}
H(x, \xi,t) = \int_\Omega \mathrm{d}\beta \mathrm{d}\alpha \,
\big[ F(\beta, \alpha,t) + \xi \, G(\beta,\alpha,t) \big] \delta(x - \beta -
\alpha \xi) \;,
\end{equation}
where $F(\beta,\alpha) \equiv F_D(\beta,\alpha)-\partial/\partial\alpha \ \chi(\beta,\alpha)$ and $G(\beta,\alpha) \equiv G_D(\beta,\alpha)-\partial/\partial\beta \  \chi(\beta,\alpha)$ and $\chi(\beta,\alpha)$ is any $\alpha$-odd function vanishing on the boundary of $\Omega$. 
The ensemble of transformations defined by all the possible functions $\chi(\beta,\alpha)$ are labelled \emph{scheme} transformations (sometimes named {\it gauge} transformations) and any of the resulting pairs $(F,G)$ constitutes a particular {\it scheme} for the DD representation of the GPD $H(x,\xi,t)$.
 In \refcite{Chouika:2017dhe}, we have generally and thoroughly discussed the three main schemes so far employed in the relevant literature, the way they are related to each other and their conditions and major implications.
 Here, let us specialize to the valence quark GPD, with support $x \in \left[-\xi,+1\right]$, and use the following representation:
%-------
\begin{eqnarray}\label{eq:DDs-P-Ds}
F(\beta,\alpha,t) &=& (1-\beta) \ h(\beta,\alpha,t) \ + \ \delta(\beta) \ D^{+}(\alpha,t) \ , \nonumber \\  
G(\beta,\alpha,t) &=& - \alpha \ h(\beta,\alpha,t) \ + \  \delta(\beta) \ D^{-}(\alpha,t) \ ,
\end{eqnarray}
where $h(\beta,\alpha,t)$ is one single function for the quark DD, with support on $\Omega^{>}=\Omega \cap \left\{\beta>0\right\}$, which fully defines the GPD within the DGLAP kinematic domain; while $D^{+}$ ($D^{-}$) is an $\alpha$-even ($\alpha$-odd) function with support $\alpha \in [-1,+1]$ which, supplemented by $\delta(\beta)$, is non-vanishing only along the line $\beta=0$, a subset of measure $0$ which only contributes to the ERBL kinematic region. 
If one plugs the DDs defined in \refeq{eq:DDs-P-Ds} into \refeq{eq:gpd-in-general-F-and-G-gauge}, the GPD would read
%------
\begin{eqnarray}
H(x, \xi,t) &=&  (1-x) \ \int_\Omega \mathrm{d}\beta \mathrm{d}\alpha \,
h(\beta,\alpha,t) \delta(x - \beta - \alpha \xi) \nonumber \\ 
&& + \; \frac 1 {|\xi |} D^{+}\left(\frac{x}{\xi},t\right) 
\; + \;  \mathrm{sgn}(\xi) \; D^{-}\left(\frac{x}{\xi},t\right)\ \;, 
\label{eq:gpd-P-Ds}
\end{eqnarray}
whence it can be easily seen that $D^{-}$ contributes to the so-called Polyakov-Weiss D-term~\cite{Polyakov:1999gs}, linked to the DD $G(\beta,\alpha,t)$ in \refeq{eq:gpd-in-general-F-and-G-gauge}, while $D^{+}$ is related to the DD $F(\beta,\alpha,t)$, and $h(\beta,\alpha,t)$ is the one single component for the DD in the {\it Pobylitsa} (P) scheme \cite{Pobylitsa:2002vi}.
 On the other hand, if one takes the forward limit in \refeq{eq:gpd-in-general-F-and-G-gauge} with the DDs given by \refeq{eq:DDs-P-Ds}, we would obtain for the PDF
%-------
\begin{eqnarray}\label{eq:conditionDp}
q(x) \ = \ H(x,0,0)  &=&  2 (1-x) \, \int_0^1 \mathrm{d}\alpha \, h(x,\alpha,0) \\ 
&+&  2 \delta(x) \, 
\underbrace{\int_0^1 \mathrm{d}\alpha \, D^{+}\left(\alpha,0\right)}_{\displaystyle =0} \ ,  
\nonumber 
\end{eqnarray} 
which makes manifest the condition to keep conserved the quark number.
 Indeed, as $D^{+}$ is required not to contribute within the DGLAP domain, then: $\int_0^1 \mathrm{d}\alpha D^{+}(\alpha,t)\equiv 0$, for any $t$, and $D^{+}$ will not contribute to the form factor either through the sum rule, 
%--------
\begin{equation}\label{eq:fpi}
F_\pi(t) \ = \ \int_{-1}^{1} \mathrm{d}x H(x,\xi,t) \ = \ \int_\Omega \mathrm{d}\beta \mathrm{d}\alpha \,
(1-\beta) h(\beta,\alpha,t) \ .
\end{equation}

Whether, and under which condition, a given GPD can be represented by DDs in the P-scheme is an issue discussed at length in \refcite{Chouika:2017dhe}. We concluded therein that a DD $h(\beta,\alpha,t)$, either summable over $\Omega$ or not, can be always obtained as a representation for any GPD.  Moreover, it was also shown in~\refcite{Chouika:2017dhe} that two GPDs with an equal DGLAP region will differ only by terms of the form $D^{+}$ and $D^{-}$, as those in \refeq{eq:gpd-P-Ds} which result from contributions to the DDs $F$ and $G$ in \refeq{eq:DDs-P-Ds} only lying along the line $\beta=0$. 

Then, eventually, the covariant extension of the GPD overlap representation will result from obtaining the DD $h(\beta,\alpha,t)$ by the inversion of \refeq{eq:gpd-P-Ds} for the DGLAP GPD and the further computation of the ERBL GPD by the direct application of the same equation with the so obtained DD.
 However, the knowledge of the GPD in the DGLAP domain can only constrain the ERBL GPD up to the additional terms given by $D^+$ and $D^-$ in \refeq{eq:gpd-P-Ds}, which generally express the ambiguity for this covariant extension. 

\section{Taming the ambiguities with the soft pion theorem}
\label{sec:soft-pion-theorem}
 
As far as $D^+$ and $D^-$ in \eqref{eq:DDs-P-Ds} have support for $\alpha$ on $[-1,1]$ and appear included in terms only defined along the line $\beta=0$, by means of $\delta(\beta)$, they solely contribute to the ERBL kinematic region, $|x| \le \xi$, as it is clearly manifest from \refeq{eq:gpd-P-Ds}.
Such terms cannot be grasped from the DGLAP information but can be, at least, constrained with the use of a soft pion theorem that states in the chiral limit~\cite{Polyakov:1998ze}:
\begin{eqnarray}
H^{I=0}\left(x,\xi=1,t=0\right)&=&0\,,\label{eq:soft_pion_theorem_isoscalar} \\
H^{I=1}\left(x,\xi=1,t=0\right)&=&\varphi_{\pi}\left(\frac{1+x}{2}\right)\,,\label{eq:soft_pion_theorem_isovector}
\end{eqnarray}
where the isoscalar (isovector) pion GPDs, $H^{I=0}$ ($H^{I=1}$), can be defined as the
odd (even) contribution to the GPD $H_{\pi^{+}}^{u}$:
\begin{eqnarray}
H^{I=0}\left(x,\xi,t\right)&=&H_{\pi^{+}}^{u}\left(x,\xi,t\right)-H_{\pi^{+}}^{u}\left(-x,\xi,t\right)\,,
\label{eq:HI0}
\\
H^{I=1}\left(x,\xi,t\right)&=&H_{\pi^{+}}^{u}\left(x,\xi,t\right)+H_{\pi^{+}}^{u}\left(-x,\xi,t\right)\,,
\label{eq:HI1}
\end{eqnarray}
and where $\varphi_{\pi}$ is the pion Distribution Amplitude (DA).

Let us define, still for the quark GPD ($-\xi \le x \le 1$), 
%------
\begin{equation}\label{eq:HP}
\HPupi(x,\xi,t) \ = \ (1-x) \, \int_\Omega \mathrm{d}\beta \mathrm{d}\alpha \, h_{\pi^+}^u(\beta,\alpha,t) \, \delta(x - \beta - \alpha \xi )
\end{equation}
where $h_{\pi^+}^u$ results from the inversion of \refeq{eq:gpd-P-Ds} with the DGLAP GPD given by the overlap of LFWFs, \refeq{eq:overlap_ud}.
 Then, after plugging \eqref{eq:gpd-P-Ds} into \eqref{eq:HI0} and the result into \eqref{eq:soft_pion_theorem_isoscalar}, one is left with: 
%------
\begin{equation}\label{eq:solDm}
D^-(x,0) = \frac 1 2 \left[ \HPupi(-x,\xi=1,t=0) \, - \, \HPupi(x,\xi=1,t=0) \right] \, \ , 
\end{equation}
which fixes the value of $D^-$ at vanishing squared momentum transfer.
 If we apply next Eqs.~(\ref{eq:gpd-P-Ds},\ref{eq:HI1}) to \eqref{eq:soft_pion_theorem_isovector}, we would have
%-------
\begin{eqnarray}
D^{+}(x,0) &=&  \frac 1 2 \left[ \varphi\left(\frac{1+x}{2}\right) \,  - \, \HPupi(x,\xi=1,t=0) \, 
\right. \nonumber \\
&& \left. - \, \HPupi(-x,\xi=1,t=0) \right] \, ,
\label{eq:solDp}
\end{eqnarray}
constraining thus $D^+$ at $t=0$. An interesting remark in order here is the following: when one performs the integration on $x$ over its support $[-1,1]$ of both sides of \eqref{eq:solDp}, the r.h.s. gives
%------
\begin{align}
&\quad \frac 1 2 \int_{-1}^1 \mathrm{d}x \, \varphi\left(\frac{1+x}{2}\right) \nonumber \\
&- \frac 1 2 \int_{-1}^1 \mathrm{d}x \left[ \HPupi(x,\xi=1,t=0) \, + \, \HPupi(-x,\xi=1,t=0) \right]  \nonumber \\ 
=&  \int_0^1 \mathrm{d}x \, \varphi(x) -  \int_{-1}^1 \mathrm{d}x \, q(x) \ = \ 0 \ ,
\end{align}
its vanishing relying only on the correct normalization of both DA and PDF and, accordingly, imposing for the l.h.s. that\footnote{The even parity of $D^+$, manifest from \refeq{eq:solDp}'s r.h.s. because $\varphi(x)$ is symmetric under the exchange $x \to 1-x$, implies 
$\int_0^1 \mathrm{d}\alpha D^+ = 0$ as the immediate consequence of its vanishing after integration over its support $[-1,1]$.} 
%------
\begin{equation}\label{eq:conditionDp-soft-pion}
\int_{0}^{1}\mathrm{d}x \, D^+(x,0) \ =\ 0 \ ,
\end{equation}
the condition given by \eqref{eq:conditionDp}, resulting here from a soft pion theorem. 

If we restrain ourselves to the pion valence-quark GPD and assume that $D^+$ is a continuous function, we can be fully general when writing
%-----
\begin{equation}\label{eq:Dp-general}
D^+(\alpha,0) \ = \ (1-\alpha^2) \sum_{i=1}^{\infty} c_i \, C^{(3/2)}_{2i}(\alpha) \ ,
\end{equation}
%-----
where the factor $1-\alpha^2$ reflects that $D(\pm 1,0)$=0, a condition imposed by factorisation, as the GPD has to be continuous at $x = \pm \xi$.
On top of this, the expansion in the orthogonal 3/2-Gegenbauer polynomials of even degree (excluding the first one, $C^{(3/2)}_0=1$) guarantees both the $\alpha$-even parity and the fulfilling of the condition \eqref{eq:conditionDp-soft-pion}, 
%-----
\begin{eqnarray}
& & \int_0^1 \mathrm{d}\alpha \, D^+(\alpha,0)   \\
&=& \frac 1 2 \sum_{i=1}^{\infty} \, c_i \, \int_{-1}^1 \mathrm{d}\alpha \, 
(1-\alpha^2) \, C^{(3/2)}_0(\alpha)\, C^{(3/2)}_{2i}(\alpha) = 0 \ . \nonumber
\end{eqnarray}
Therefore, $D^+$ and $D^-$ can be always chosen so that the soft pion theorem expressed by \refeqs{eq:soft_pion_theorem_isoscalar}{eq:soft_pion_theorem_isovector} may be fulfilled and, for the same price, the ambiguities in the covariant extension from DGLAP to ERBL domains be constrained at vanishing squared momentum transfer. 

Indeed, the issue of the observance of the soft pion theorem can be approached in the other way around.

 We should emphasise once more that, in terms of LFWFs, the ERBL region is understood as an overlap of $N$ and $N+2$ partons LFWFs, starting in the case of the pion at $N=2$. 
 On the other hand, the covariant extension based on the Radon transform insures the polynomiality property, and any idea of Fock state truncation in the ERBL region is lost.
 One can only say that the information from higher Fock states LFWFs required to fulfil polynomiality is properly captured.
 But since the PDA is completely described by the two-body LFWF, one can wonder whether there is some genuine information in the 4-body LFWF interplaying with the 2-body one via overlap to produce a GPD fulfilling the soft pion theorem in our lowest-Fock-states approach.

 Rephrasing the question in a more technical way, in connection with the Radon transform representation: does the information along the line $\beta=0$ in DD space play a crucial role to guarantee the correct limit in the ERBL maximally skewed kinematic?
 To the extent of our knowledge, there is no conclusive answer to this question.
 Previous results~\cite{Mezrag:2014jka} have shown how critical the implementation of the Axial-Vector Ward-Takahashi identity is when solving the Dyson-Schwinger and Bethe-Salpeter equations in order to fulfil the soft pion theorem in covariant computations.
 We certainly expect the same thing to be true within the overlap of LFWFs framework.
 If the covariant extension of the DGLAP GPD obtained from the appropriate 2-body LFWFs is not sufficient to fully reconstruct the ERBL in the kinematic limit of the soft pion theorem, the terms $D^+$ and $D^-$ {\it should} be eventually adjusted in any case to supply a full description of the pion.  

\section{The Nakanishi-based Bethe-Salpeter model}
\label{sec:AlgebraicModel}

We have described a systematic and fully general prescription aimed at obtaining a hadron GPD, on its entire kinematic domain, from the knowledge of the relevant LFWFs.
 The prescription is essentially based on accommodating the overlap of these LFWFs within the DD representation.
 The pion has been so far used as a simple guiding case and, still in what follows, we will consider a specific pion GPD model, the one introduced in \refcite{Mezrag:2016hnp} in order to illustrate this prescription. 
 Owing to the simplicity of the pion model, we will produce fully algebraic results and, very specially, show how to deal with the soft pion theorem and the ambiguities in the covariant extension from DGLAP to ERBL kinematic domains. 

\begin{figure}[t]
%--
\centerline{\includegraphics[width=0.9\linewidth]{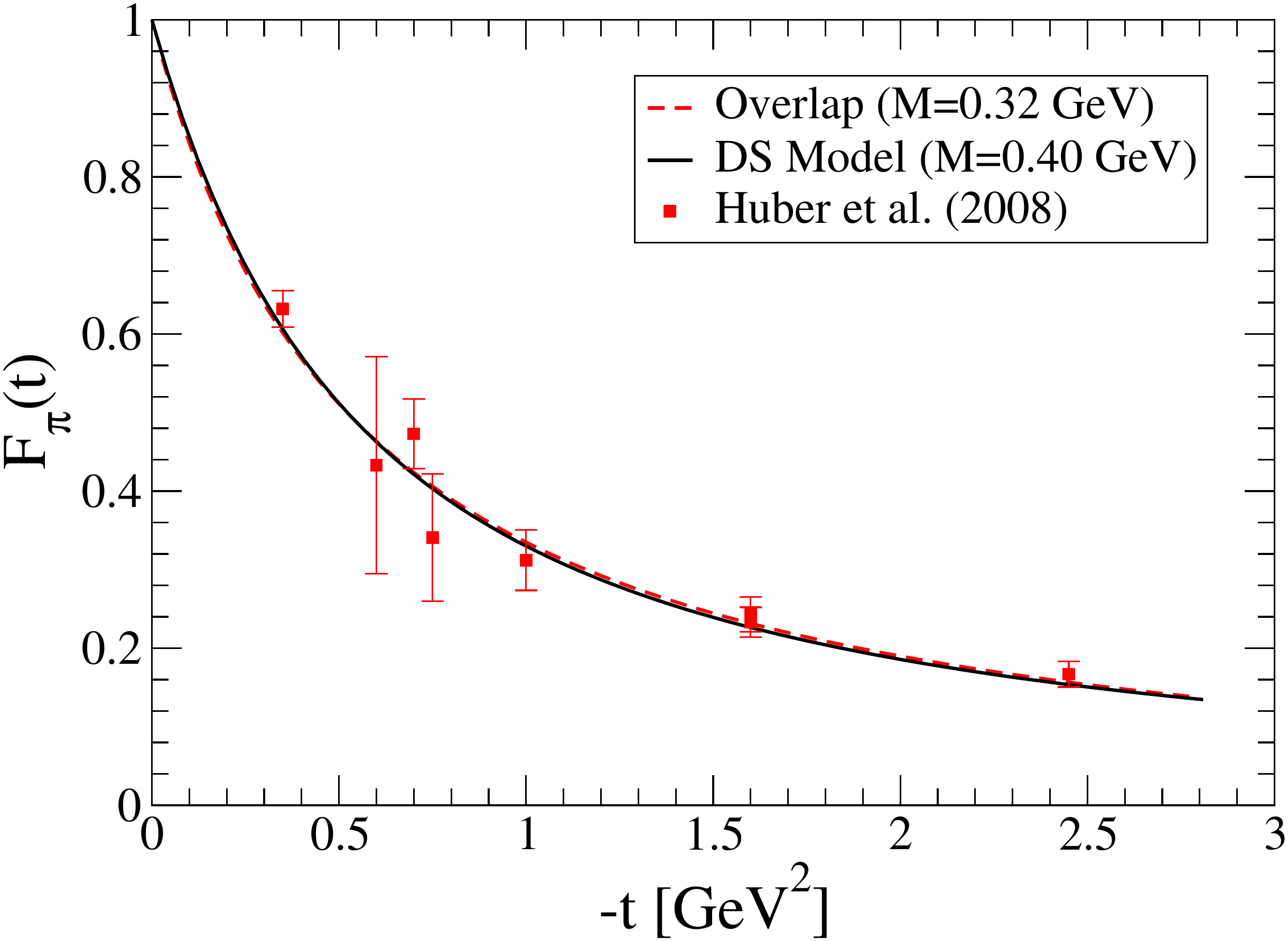}}
%--
\caption{Pion's electromagnetic form factor expressed by \refeq{eq:fpiH} (dashed red line), with the model mass parameter fixed by \refeq{eq:valM} and displayed in terms of $-t$ in GeV, compared to experimental data~\cite{Huber:2008id} (red solid circles) and to the results obtained in \refcite{Mezrag:2014jka} within a covariant DS-inspired calculation (black solid line) where a different model mass parameter, $M=0.40$ GeV, was determined. 
\label{fig:fpi}
}
%--
\end{figure}

The basic ingredient for the GPD construction is the LFWF obtained by the appropriate integration and projection of the pion Bethe-Salpeter wave function resulting from the algebraic model described in \cite{Chang:2013pq} and based on its Nakanishi representation~\cite{Nakanishi:1963zz,Nakanishi:1969ph}.
In this model developed in euclidean space, the quark propagator is $S(q)=[-i\gamma \cdot q + M]/[q^2+M^2]$ and the Bethe-Salpeter amplitude is given by:
\begin{equation}
  \label{eq:BSA}
  \Gamma_\pi(q,P) = i \mathcal{N}\gamma_5 \int_0^\infty \textrm{d}\omega \int_{-1}^{1} \textrm{d}z \frac{\rho(\omega,z)M^2}{\left(q- \frac{1-z}{2}P \right)^2 + M^2+\omega },
\end{equation}
where $\rho(\omega,z)$ is the Nakanishi weight modelled as $\rho(\omega,z) =\delta(\omega) (1-z^2)$ and $\mathcal{N}$ is an overall normalization constant.
The Bethe-Salpeter wave function is obtained as $S(q)\Gamma_\pi(q,P)S(q-P)$.
As shown in~\refcite{Mezrag:2016hnp} (the details of the computation can be found therein), there are two contributions to the LFWF, the helicity-0:
%--------
\begin{equation}\label{eq:LCWF}
\Psi_{l=0}\left(x,\vperp{k}\right) = 8 \, \sqrt{15} \, \pi \, \frac{M^3}{\left(\vperp{k}^2 + M^2\right)^2} \, \left(1-x\right) x \, , 
\end{equation}
%--------
and the helicity-1: 
%--------
\begin{equation}\label{eq:LCWFh1}
i \, {k_{\perp}}_j \, \Psi_{l=1}\left(x,\vperp{k}\right) = 8 \, \sqrt{15} \, \pi \, \frac{{k_{\perp}}_j \, M^2}{\left(\vperp{k}^2 + M^2\right)^2} \, \left(1-x\right) x \, 
\end{equation}
%--------
with $j=1,2$ and where $M$ is the model mass parameter introduced above. 
One can readily notice that our LFWFs do not show any $(x,k_\perp)$ correlations, and can be written as $f(x) \, g(\vperp{k}^2)$. This is due to our simple choice of the Nakanishi weight in \refeq{eq:BSA}, $\rho(\omega,z)$. There is no doubt that correlations would arise from a proper solution of the BSE. But, despite the lack of correlations, this simple algebraic model remains insightful for our exploratory work, and illustrates well our extension technique.
Proceeding with the computations, Eqs.~(\ref{eq:LCWF},\ref{eq:LCWFh1}) can then be combined into the following expression, 
\begin{eqnarray}
\label{eq:overlap_ud_2contributions}
&& \left. H_{\pi^{+}}^{u}\left(x,\xi,t\right)\right|_{\xi \le x} =  \\
&& \int\frac{\mathrm{d}^{2}\mathbf{k}_{\bot}}{16\,\pi^{3}} \; \left[
\Psi_{l=0}^{*}\left(\frac{x-\xi}{1-\xi},\hat{\mathbf{k}}_{\bot} \right) 
\Psi_{l=0}\left(\frac{x+\xi}{1+\xi},\tilde{\mathbf{k}}_{\bot}\right) \right. \nonumber \\
&& + \left.  \hat{\mathbf{k}}_{\bot} \cdot \tilde{\mathbf{k}}_{\bot}  \; 
 \Psi_{l=1}^{*}\left(\frac{x-\xi}{1-\xi},\hat{\mathbf{k}}_{\bot} \right)
 \Psi_{l=1}\left(\frac{x+\xi}{1+\xi}, \tilde{\mathbf{k}}_{\bot} \right)
 \right]\,,  \nonumber 
\end{eqnarray}
%-----------------
with $\hat{\mathbf{k}}_{\bot}=\mathbf{k}_{\bot}+\frac{1-x}{1-\xi} \frac{\Delta_\bot}{2}$ and 
$\tilde{\mathbf{k}}_{\bot}=\mathbf{k}_{\bot}-\frac{1-x}{1+\xi} \frac{\Delta_\bot}{2}$,  
which extends \refeq{eq:overlap_ud} for the GPD of our special $\pi^+$ case.
One is thus left with:
%----------
\begin{eqnarray}
\label{eq:GPDAlgebraicDGLAP}
\left. H_{\pi^{+}}^{u}\left(x,\xi,t\right)\right|_{\xi \le x} \; = \;
\frac{15}{2} \ \frac{(1-x)^2 (x^2-\xi^2)}{(1-\xi^2)^2} \ \frac 1 {(1 + \zeta)^2} \\ 
\left( 3 + \frac{1-2 \zeta}{1+\zeta} \frac{\displaystyle 
\arctanh{\left(\sqrt{ \frac{\zeta}{1+\zeta}}\right)}}{ \displaystyle \sqrt{\frac{\zeta}{1+\zeta}}} \right) \ ,
\nonumber 
\end{eqnarray}
as a fully algebraic result for the DGLAP region, where
\begin{equation}\label{eq:x-xi}
\zeta \ = \ \frac{-t}{4M^2} \frac{(1-x)^2}{1-\xi^2} \;, 
\end{equation}
%--------
encodes the correlated dependence of the kinematical variables $x$ and $t$, as a natural translation of the kinematical structure of Eqs.~(\ref{eq:LCWF},\ref{eq:LCWFh1}).
%----
 It should be noticed that such a correlation is fully consistent with the results of pQCD when $x\rightarrow 1^-$, as any $t$-dependence in \refeq{eq:GPDAlgebraicDGLAP} appears thus suppressed by a factor $(1-x)^2$~\cite{Yuan:2003fs}.
 Indeed, in this limit, \refeq{eq:GPDAlgebraicDGLAP} yields:
 \begin{equation}
   \label{eq:LargexResults}
   \left. H_{\pi^+}^{u}(x,\xi,t)\right|_{\xi \le x} =  30 \frac{(1-x)^2}{1-\xi^2} \left( 1 \, - 2 \frac{1-x}{1-\xi^2} \, 
   + \, \mathcal{O}\left( (1-x)^2\right) \right)  \, ,
 \end{equation}
 %----
where the leading term plainly agrees with the one obtained in~\refcite{Yuan:2003fs}\footnote{The limit $x \rightarrow 1^-$ of \eqref{eq:GPDAlgebraicDGLAP} given by \refeq{eq:LargexResults} is equivalent to $q(x)/(1-\xi^2)$, as it is displayed by Eq.(4) of \refcite{Yuan:2003fs}.}, while the first subleading correction is also shown not to depend on $t$.
The forward limit of \eqref{eq:GPDAlgebraicDGLAP},
%-----
\begin{equation}
q(x) = H_{\pi^+}^u(x,0,0) = 30 x^2 (1-x)^2 \, ,
\end{equation}
%-----
yields the same result which is found in \refcite{Mezrag:2014jka} as an excellent approximation for the pion's valence dressed-quark PDF~\cite{Chang:2014lva}.
 Furthermore, applying the sum rule for the electromagnetic pion form factor, expressed by \eqref{eq:fpi}, one is left with

%-----
\begin{align}
F_\pi(t) = &~ 720 \, \frac {M^4}{t^2} \ \left( 1 - \sqrt{\frac{4-t/M^2}{-t/M^2}} \, \arctanh{\left(\sqrt{\frac{-t/M^2}{4-t/M^2}}\right)} \right.  \nonumber \\ \label{eq:fpiH}
&~ \left. + \, \frac 1 3 \,  \arctanh^2{\left(\sqrt{\frac{-t/M^2}{4-t/M^2}}\right)} \, \right) \\
= &~ 1\, - \, \frac{4}{21} \left(-\frac t {M^2}\right) \, + \, {\cal O}(t^2) \ ,
\end{align}  
%------
whence the model mass parameter can be identified as
%------
\begin{equation}\label{eq:valM}
M \ = \sqrt{\frac{24}{21}} \, \frac 1 {r_\pi} \ = \ 318 \pm 4 \ \mbox{\rm MeV} \ , 
\end{equation}
%------
where we use $F_\pi(t) \simeq 1- r_\pi^2/6 \, (-t)$ and take for the pion electric charge radius: $r_\pi=0.672 \pm 0.008$~fm~\cite{Beringer:1900zz}.
 Thus, \refeq{eq:fpiH} supplemented with \refeq{eq:valM} provide us with a model prediction which, as can be seen in~\reffig{fig:fpi},  compare fairly well with contemporary data, up to $-t \simeq 2.5$ GeV$^2$.
 At large $t$, nonetheless, \refeq{eq:fpiH} behaves as $1/t^2$, whereas the expected behavior is $1/t$~\cite{Efremov:1978rn,Lepage:1980fj}.
 This wrong behaviour can be well understood, as explained in \refcite{Mezrag:2014jka}, because \refeqs{eq:LCWF}{eq:LCWFh1} have been derived from a Bethe-Salpeter wave function omitting contributions from the pseudovector components that are required for a complete description of the pion~\cite{Maris:1997hd,Qin:2014vya}.
 One should also keep in mind that, in the covariant approach, the large $t$ behaviour can also be produced by the dressing of the insertion \cite{deMelo:2005cy,Frederico:2009fk}.

\begin{figure}[t] 
%--
\begin{tabular}{cc}
\leftline{\includegraphics[width=0.65\linewidth]{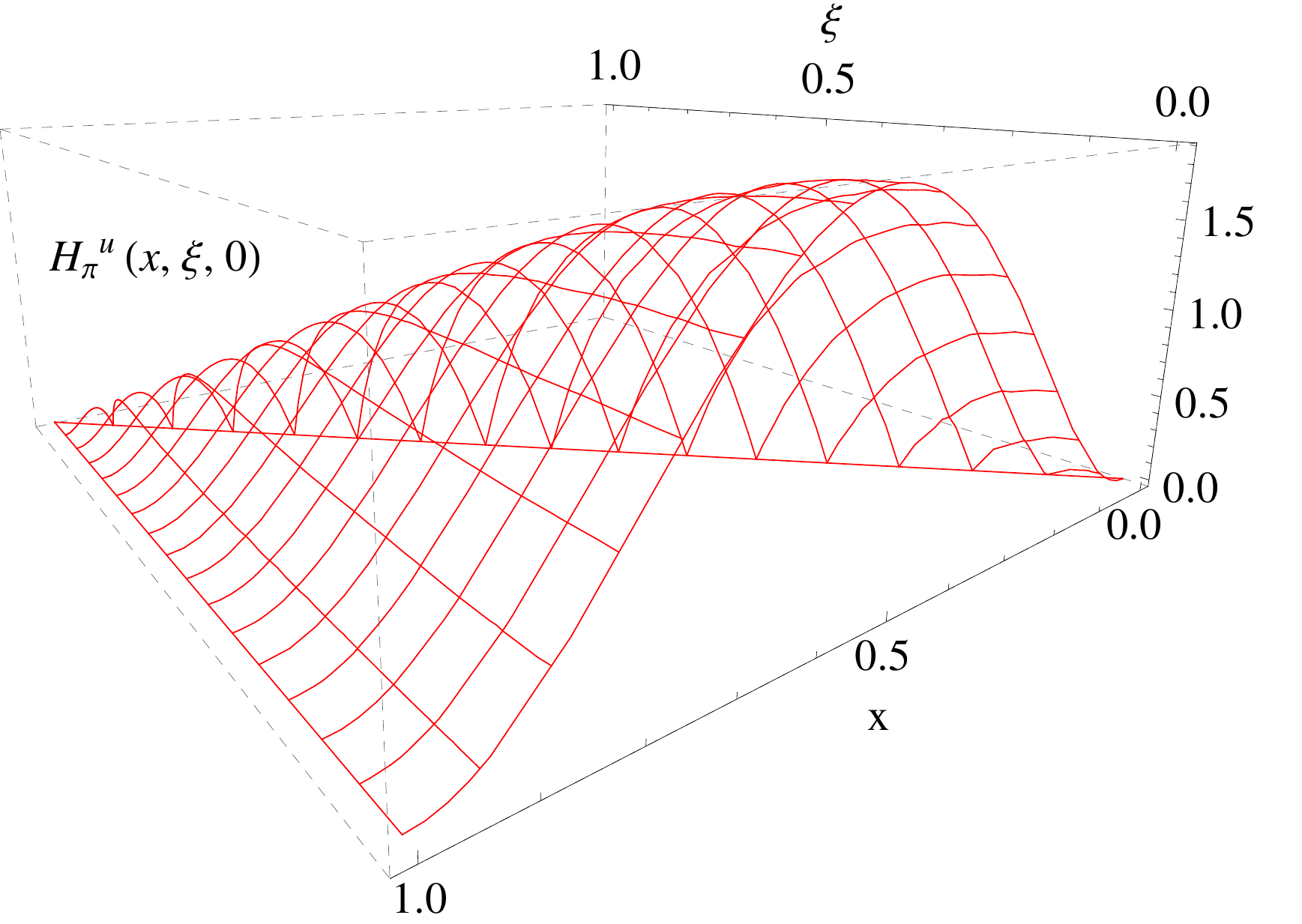}} 
\vspace*{-5ex}
\\
\rightline{\includegraphics[width=0.65\linewidth]{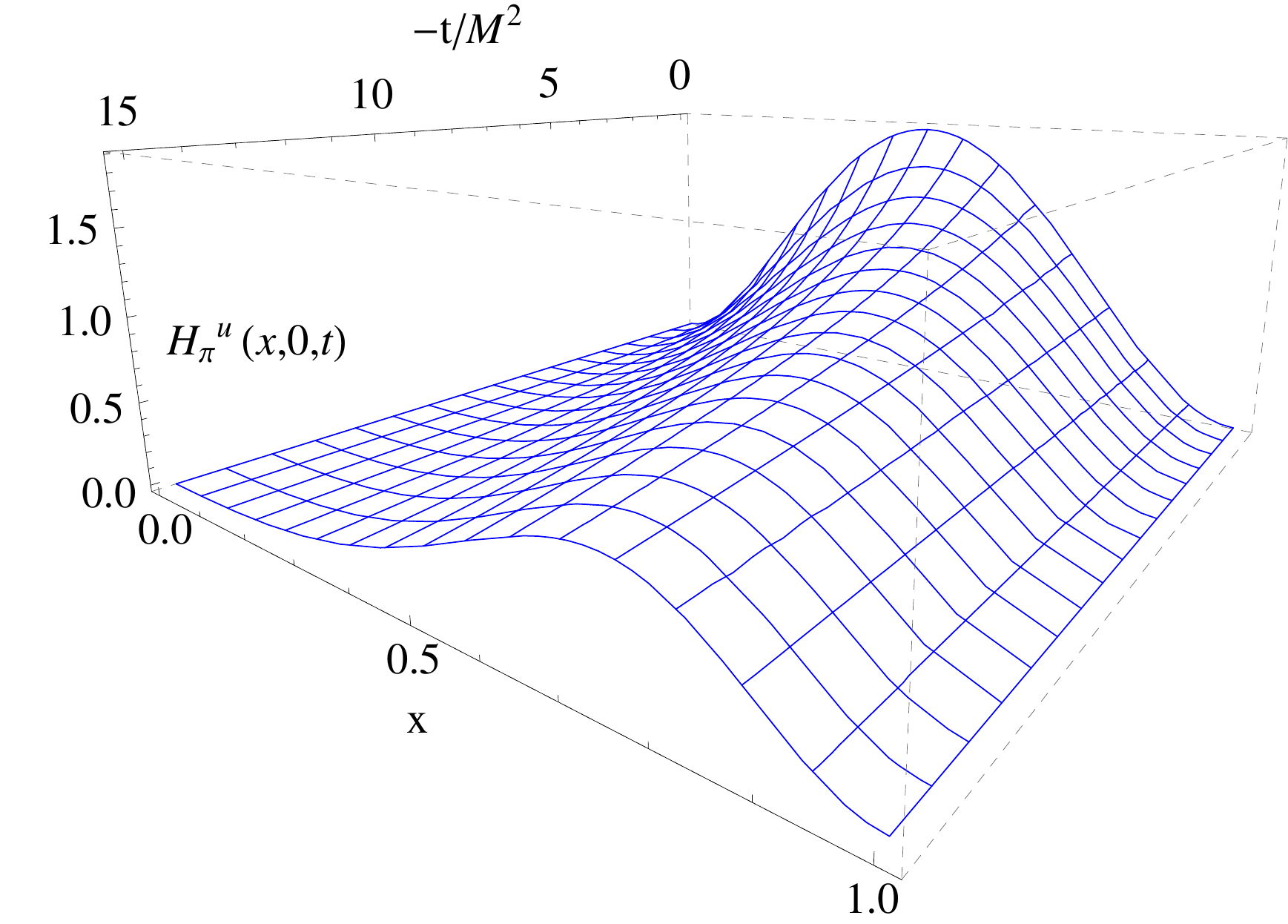}}
\end{tabular}
%--
\caption{DGLAP GPD given by \refeq{eq:GPDAlgebraicDGLAP} and obtained from the overlap of LFWFs, \refeqs{eq:LCWF}{eq:LCWFh1}, expressed by \refeq{eq:overlap_ud_2contributions}, plotted in terms of $x$ and $\xi$, at $t=0$ (upper panel); and in terms of $x$ and $t$ at $\xi=0$ (lower panel). 
\label{fig:GPD-DGLAP}}
%-- 
\end{figure}

Then, as well the PDF as the pion form factor that result from \refeq{eq:GPDAlgebraicDGLAP} consistently agree with the zero skewness GPD sketched in \refcite{Mezrag:2014jka}, within the context of a covariant calculation inspired by the solutions of Dyson-Schwinger (DS) and Bethe-Salpeter (BS) equations (see Fig.~\ref{fig:fpi}). 
More than this, \refeq{eq:GPDAlgebraicDGLAP} specialized at $\xi=0$ can be readily accommodated within the general form given by Eq.~(16a) in \refcite{Mezrag:2014jka}, 
%-----
\begin{equation}
H_{\pi^+}^u(x,0,t) = q(x) \mathcal{N}(t) C_\pi(x,t) F_\pi(t) \, 
\end{equation}
%-----
such that the function $C_\pi(x,t)$, defined to express the $(x,t)$ correlations in the GPD, takes the form
%----    
 \begin{equation}
 C_\pi(x,t) \, = \, C_\pi^{DS}(x,t) \, \left( 1 -  \frac 2 3 \zeta_0 + \mathcal{O}(\zeta_0^2) \right) 
 \end{equation}
%----
with $C_\pi^{DS}(x,t)=1/(1+\zeta_0)^2$ being the result obtained in \refcite{Mezrag:2014jka} and $\zeta_0=\zeta(\xi=0)=-t/[4M^2](1-x)^2$, while 
%-----
\begin{equation}
\frac{1}{\mathcal{N}(t)} =  \int_{-1}^{1} \mathrm{d}x \, q(x) C_\pi(x,t)  =  \frac 1 {\mathcal{N}^{DS}(t)}  
\left( 1 - \frac{1}{21} \, \frac{-t}{M^2} + \mathcal{O}(t^2) \right) \; . 
\end{equation}
%----
Then, at low-$t$, $C_\pi(x,t) \simeq C_\pi^{DS}(x,t)$ and $\mathcal{N}(t) \simeq \mathcal{N}^{DS}(t)$ such that  \refeq{eq:GPDAlgebraicDGLAP} can both support the approximations made in \refcite{Mezrag:2014jka} and be understood as an extension, beyond the zero skewness limit, of the results therein obtained. This extended DGLAP GPD appears displayed in~\reffig{fig:GPD-DGLAP}.     
 
Now, according to the prescription described in the previous section, the first step for the covariant extension from DGLAP to ERBL domains of \refeq{eq:GPDAlgebraicDGLAP} consists in performing the inversion of the Radon transform in \refeq{eq:gpd-P-Ds} for the DGLAP GPD, and obtaining thus the DD in the P scheme.
 A careful computation, based on a sensible choice of trial functions, allows for the derivation of the 
following closed expression:
\begin{align}
h_{\pi^+}^u&(\beta, \alpha, t) = \frac {15}{2}  \theta(\beta) \, \left[1 + \frac{-t}{4 M^2} \left(\left(1-\beta\right)^2-\alpha^2\right)\right]^{-3} \nonumber \\
&\times \, \left[ 1 -3 (\alpha^2-\beta^2) - 2 \beta + \frac{-t}{4 M^2} \left(1 - (\alpha^2- \beta^2)^2 - 4\beta(1-\beta) \right) \right] \, ,
 \label{eq:DDAlgebraic}
\end{align}
which, plugged into \refeq{eq:HP}, gives
%------
\begin{equation}
\label{eq:GPDAlgebraicERBL}
\left. {H_P}_{\pi^+}^{u}(x,\xi,0)\right|_{|x| \le \xi} \ = \
\frac {15} 2 \frac{(1-x) (\xi^2-x^2)}{\xi^3 (1+\xi)^2} \left( x + 2 x \xi + \xi^2 \right) \ ,
\end{equation}
%-------
a simple closed expression for the ERBL GPD in the case \mbox{$t=0$}.
 Of course, \refeq{eq:HP} can also provide us with ERBL GPD results for any nonvanishing $t$.
 We will however focus on \mbox{$t=0$}, wherein, for the pion's case and as explained in~\refsec{sec:soft-pion-theorem}, there is a unique way to perform the covariant extension by fulfilling the soft pion theorem. 
\begin{figure}[t]
%--
\centerline{\includegraphics[width=0.9\linewidth]{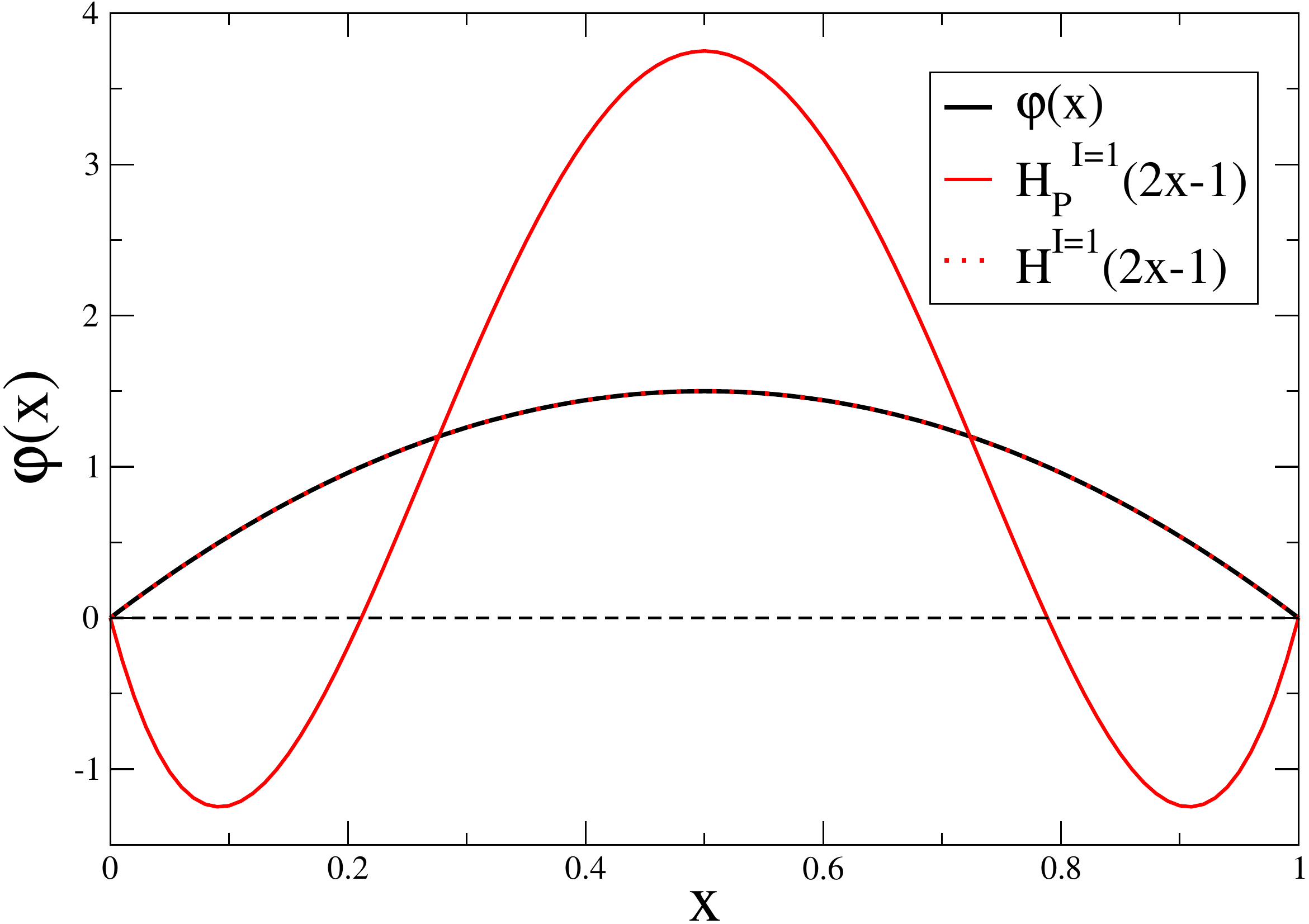}}
%--
\caption{Check of the soft pion theorem.
$H_P^{I=1}(2x-1,1)$ (red solid line), the maximally skewed GPD with $D^+(\alpha,0)\equiv 0$, clearly differs from the asymptotic pion's distribution amplitude, $\varphi(x)$ (black solid line).
Only after incorporating the term $D^+(\alpha,0)$ given by \eqref{eq:solDp-algebraic} do $H^{I=1}(2x-1,1)$ (red dotted line) and $\varphi(x)$ agree with each other, as dictated by \refeq{eq:soft_pion_theorem_isovector}.  
\label{fig:soft-pion}
}
%--
\end{figure}

Indeed, if one applies \refeq{eq:GPDAlgebraicERBL} to \refeq{eq:solDm}, the additional D-term is constrained by
%----
\begin{equation}\label{eq:solDm-algebraic}
D^-(\alpha,0) \, = \,  - \frac{15}{4} \, \alpha (1-\alpha^2) \ , 
\end{equation} 
%----
such that \eqref{eq:soft_pion_theorem_isoscalar} is observed.
In addition, the asymptotic DA, $\varphi(x) = 6 (1-x) x$ can be straightforwardly derived from the LFWFs\footnote{The asymptotic DA can be also directly obtained from the BSA, \refeq{eq:BSA}, as shown in \refcite{Chang:2013pq}.} expressed by \refeqs{eq:LCWF}{eq:LCWFh1} and, together with the ERBL GPD in~\refeq{eq:GPDAlgebraicERBL}, plugged into \refeq{eq:solDp} to give: 
%----
\begin{equation}\label{eq:solDp-algebraic}
D^+(\alpha,0) \, = \, \frac 9 8 \, (1-\alpha^2) (5\alpha^2-1) \, = \, \frac 3 4 \, (1-\alpha^2) \, C^{(3/2)}_2(\alpha) \ , 
\end{equation}
%----
which, in particular, corresponds to the general form given by \refeq{eq:Dp-general}, with $c_2=3/4$ and, otherwise, $c_i=0$. 

\begin{figure}[t]
%--
\centerline{\includegraphics[width=0.95\linewidth]{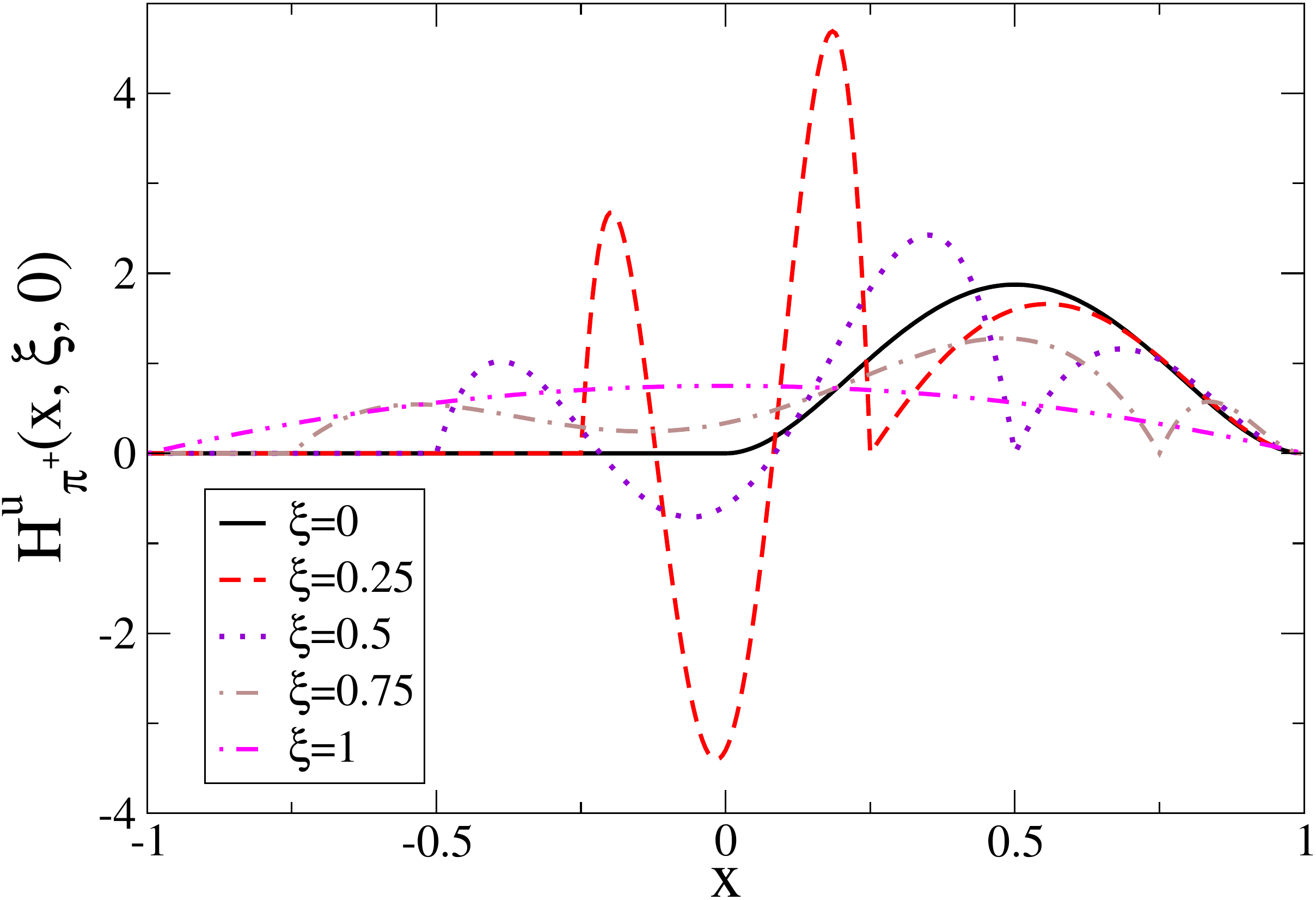}}
%--
\caption{The full GPD at $t=0$, expressed by \refeq{eq:gpd-P-Ds}, where ${H_P}_{\pi^+}^u$ is given by \refeq{eq:GPDAlgebraicDGLAP} for DGLAP and by \refeq{eq:GPDAlgebraicERBL} for ERBL, and $D^-$ and $D^+$ appear constrained by \refeq{eq:solDm-algebraic} and \refeq{eq:solDp-algebraic}, respectively. The results displayed stand for $\xi=0,0.25,0.5,0.75$ and $1$. 
\label{fig:full-GPD}
}
%--
\end{figure}

Then, as can be seen in Fig.~\ref{fig:soft-pion}, only when \refeq{eq:GPDAlgebraicERBL} is supplemented by \refeqs{eq:solDm-algebraic}{eq:solDp-algebraic}, as indicated by \refeq{eq:gpd-P-Ds}, the full GPD fulfills the soft pion theorem, \refeqs{eq:soft_pion_theorem_isoscalar}{eq:soft_pion_theorem_isovector}.
This full GPD appear displayed in Fig.~\ref{fig:full-GPD}, as a function of $x$, at $t=0$ and for $\xi=0,0.25,0.5,0.75$ and $1$. It is worthwhile to notice that the oscillatory behaviour displayed by the ERBL GPD, the more and more manifest when $\xi \rightarrow 0$, results from the structure of the term $D^+$, generally written in \refeq{eq:Dp-general}, and that can be in no way inferred from the knowledge of the GPD within the DGLAP kinematic domain. 

\section{Discussion and Conclusions \label{sec:conclusion}}

The systematic technique developed very recently in \refcite{Chouika:2017dhe} for a GPD model building based on the knowledge of the hadron LFWFs, their overlap representation and the inverse Radon transform approach, thus respecting both polynomiality and positivity at the same footing, has been here illustrated by being applied to a particular simple case: a pion's valence-quark GPD model constructed on the basis of a LFWF derived from a pion's BSA built within the Nakanishi representation. 
The kinematical structure of the LFWFs remains simple, as it results from a BSA that disregards some relevant contributions for a complete pion's description.
As a consequence, a realistic nonperturbative PDA or the correct large-$t$ power behaviour for the pion's form factor remains for instance out of the model's scope. 
However, owing to this simplicity, fully algebraic closed results have been obtained for any kinematics; and being so, the model has revealed itself to be very insightful, yielding explicit correlations among the GPDs variables beyond the usual Regge parametrizations, in agreement with the predictions of pQCD.
In particular, the model predicts the pion's form factor in fair agreement with empiric information up to $-t \simeq 2.5$ GeV, with the pion's electric charge radius as the only input, and yields a PDF in the forward limit and a zero skewness GPD at low-$t$ both in excellent agreement with the results obtained within the DS- and BS-inspired covariant approach in \refcite{Mezrag:2014jka}.
As the well-reproduced leading contribution, in this latter case, is independent of $t$, the commonplace for all the model's achievements is that the large-$t$ kinematical region appears not to be under consideration.
This seems to suggest that the use of more sophisticated LFWFs models will essentially impact the large-$t$ region.
One should anyhow keep in mind that the Nakanishi representation, is completely general.
The very same procedure can thus be applied with more realistic BSA and propagators including DCSB effects like the running quarks and gluons masses.
This work therefore paves the way for a proper evaluation of the contribution of the leading Fock state to the 3D structure of the pion and, beyond, of the one of the nucleon.

Last but not least, in addition to illustrating the approach of \refcite{Chouika:2017dhe} by building a simple but fruitful algebraic pion GPD model, we have also shown how a soft-pion theorem can be invoked to constrain the ambiguities which result from the covariant extension from  DGLAP to ERBL.
As far as the theorem relies on the chiral symmetry and Ward-Takahashi identities, a tantalizing connection between underlying  symmetries and a univocal relation of the GPD descriptions within DGLAP and ERBL domains appears also to be herefrom suggested, at least in the pion case.   

\section*{Acknowledgments}
\label{sec:acknowledgments}

The authors would like to thank 
J-F. Mathiot, C.D.~Roberts, G. Salmè and C. Shi 
for valuable discussions and comments.
This work is partly supported by the Commissariat à l’Energie Atomique et aux Energies Alternatives, 
the GDR QCD “Chromodynamique Quantique”, the ANR-12-MONU-0008-01 “PARTONS”
and the Spanish ministry Research Project FPA2014-53631-C2-2-P.

\section*{References}

\bibliographystyle{model1a-num-names}
\bibliography{Bibliography}

\begin{thebibliography}{52}
\expandafter\ifx\csname natexlab\endcsname\relax\def\natexlab#1{#1}\fi
\providecommand{\url}[1]{\texttt{#1}}
\providecommand{\href}[2]{#2}
\providecommand{\path}[1]{#1}
\providecommand{\DOIprefix}{doi:}
\providecommand{\ArXivprefix}{arXiv:}
\providecommand{\URLprefix}{URL: }
\providecommand{\Pubmedprefix}{pmid:}
\providecommand{\doi}[1]{\href{http://dx.doi.org/#1}{\path{#1}}}
\providecommand{\Pubmed}[1]{\href{pmid:#1}{\path{#1}}}
\providecommand{\bibinfo}[2]{#2}
\ifx\xfnm\relax \def\xfnm[#1]{\unskip,\space#1}\fi
%Type = Article
\bibitem[{Burkardt(2000)}]{Burkardt:2000za}
\bibinfo{author}{M.~Burkardt}, \bibinfo{journal}{Phys. Rev.}
  \bibinfo{volume}{D62} (\bibinfo{year}{2000}) \bibinfo{pages}{071503}.
  \DOIprefix\doi{10.1103/PhysRevD.62.071503, 10.1103/PhysRevD.66.119903}.
  \href{http://arxiv.org/abs/hep-ph/0005108}{\tt arXiv:hep-ph/0005108},
  \bibinfo{note}{[Erratum: Phys. Rev.D66,119903(2002)]}.
%Type = Article
\bibitem[{Mueller et~al.(1994)Mueller, Robaschik, Geyer, Dittes, and
  Ho\v{r}e\v{j}si}]{Mueller:1998fv}
\bibinfo{author}{D.~Mueller}, \bibinfo{author}{D.~Robaschik},
  \bibinfo{author}{B.~Geyer}, \bibinfo{author}{F.~Dittes},
  \bibinfo{author}{J.~Ho\v{r}e\v{j}si}, \bibinfo{journal}{Fortsch.Phys.}
  \bibinfo{volume}{42} (\bibinfo{year}{1994}) \bibinfo{pages}{101--141}.
  \DOIprefix\doi{10.1002/prop.2190420202}.
  \href{http://arxiv.org/abs/hep-ph/9812448}{\tt arXiv:hep-ph/9812448}.
%Type = Article
\bibitem[{Ji(1997)}]{Ji:1996nm}
\bibinfo{author}{X.-D. Ji}, \bibinfo{journal}{Phys.Rev.} \bibinfo{volume}{D55}
  (\bibinfo{year}{1997}) \bibinfo{pages}{7114--7125}.
  \DOIprefix\doi{10.1103/PhysRevD.55.7114}.
  \href{http://arxiv.org/abs/hep-ph/9609381}{\tt arXiv:hep-ph/9609381}.
%Type = Article
\bibitem[{Radyushkin(1997)}]{Radyushkin:1997ki}
\bibinfo{author}{A.~Radyushkin}, \bibinfo{journal}{Phys.Rev.}
  \bibinfo{volume}{D56} (\bibinfo{year}{1997}) \bibinfo{pages}{5524--5557}.
  \DOIprefix\doi{10.1103/PhysRevD.56.5524}.
  \href{http://arxiv.org/abs/hep-ph/9704207}{\tt arXiv:hep-ph/9704207}.
%Type = Article
\bibitem[{Ji(1998)}]{Ji:1998pc}
\bibinfo{author}{X.-D. Ji}, \bibinfo{journal}{J.Phys.} \bibinfo{volume}{G24}
  (\bibinfo{year}{1998}) \bibinfo{pages}{1181--1205}.
  \DOIprefix\doi{10.1088/0954-3899/24/7/002}.
  \href{http://arxiv.org/abs/hep-ph/9807358}{\tt arXiv:hep-ph/9807358}.
%Type = Article
\bibitem[{Goeke et~al.(2001)Goeke, Polyakov, and Vanderhaeghen}]{Goeke:2001tz}
\bibinfo{author}{K.~Goeke}, \bibinfo{author}{M.~V. Polyakov},
  \bibinfo{author}{M.~Vanderhaeghen}, \bibinfo{journal}{Prog.Part.Nucl.Phys.}
  \bibinfo{volume}{47} (\bibinfo{year}{2001}) \bibinfo{pages}{401--515}.
  \DOIprefix\doi{10.1016/S0146-6410(01)00158-2}.
  \href{http://arxiv.org/abs/hep-ph/0106012}{\tt arXiv:hep-ph/0106012}.
%Type = Article
\bibitem[{Diehl(2003)}]{Diehl:2003ny}
\bibinfo{author}{M.~Diehl}, \bibinfo{journal}{Phys.Rept.} \bibinfo{volume}{388}
  (\bibinfo{year}{2003}) \bibinfo{pages}{41--277}.
  \DOIprefix\doi{10.1016/j.physrep.2003.08.002}.
  \href{http://arxiv.org/abs/hep-ph/0307382}{\tt arXiv:hep-ph/0307382}.
%Type = Article
\bibitem[{Belitsky and Radyushkin(2005)}]{Belitsky:2005qn}
\bibinfo{author}{A.~Belitsky}, \bibinfo{author}{A.~Radyushkin},
  \bibinfo{journal}{Phys.Rept.} \bibinfo{volume}{418} (\bibinfo{year}{2005})
  \bibinfo{pages}{1--387}. \DOIprefix\doi{10.1016/j.physrep.2005.06.002}.
  \href{http://arxiv.org/abs/hep-ph/0504030}{\tt arXiv:hep-ph/0504030}.
%Type = Article
\bibitem[{Boffi and Pasquini(2007)}]{Boffi:2007yc}
\bibinfo{author}{S.~Boffi}, \bibinfo{author}{B.~Pasquini},
  \bibinfo{journal}{Riv.Nuovo Cim.} \bibinfo{volume}{30} (\bibinfo{year}{2007})
  \bibinfo{pages}{387}. \DOIprefix\doi{10.1393/ncr/i2007-10025-7}.
  \href{http://arxiv.org/abs/0711.2625}{\tt arXiv:0711.2625}.
%Type = Article
\bibitem[{Guidal et~al.(2013)Guidal, Moutarde, and
  Vanderhaeghen}]{Guidal:2013rya}
\bibinfo{author}{M.~Guidal}, \bibinfo{author}{H.~Moutarde},
  \bibinfo{author}{M.~Vanderhaeghen}, \bibinfo{journal}{Rept.Prog.Phys.}
  \bibinfo{volume}{76} (\bibinfo{year}{2013}) \bibinfo{pages}{066202}.
  \DOIprefix\doi{10.1088/0034-4885/76/6/066202}.
  \href{http://arxiv.org/abs/1303.6600}{\tt arXiv:1303.6600}.
%Type = Article
\bibitem[{Mueller(2014)}]{Mueller:2014hsa}
\bibinfo{author}{D.~Mueller}, \bibinfo{journal}{Few Body Syst.}
  \bibinfo{volume}{55} (\bibinfo{year}{2014}) \bibinfo{pages}{317--337}.
  \DOIprefix\doi{10.1007/s00601-014-0894-3}.
  \href{http://arxiv.org/abs/1405.2817}{\tt arXiv:1405.2817}.
%Type = Article
\bibitem[{Guidal et~al.(2005)Guidal, Polyakov, Radyushkin, and
  Vanderhaeghen}]{Guidal:2004nd}
\bibinfo{author}{M.~Guidal}, \bibinfo{author}{M.~Polyakov},
  \bibinfo{author}{A.~Radyushkin}, \bibinfo{author}{M.~Vanderhaeghen},
  \bibinfo{journal}{Phys.Rev.} \bibinfo{volume}{D72} (\bibinfo{year}{2005})
  \bibinfo{pages}{054013}. \DOIprefix\doi{10.1103/PhysRevD.72.054013}.
  \href{http://arxiv.org/abs/hep-ph/0410251}{\tt arXiv:hep-ph/0410251}.
%Type = Article
\bibitem[{Goloskokov and Kroll(2005)}]{Goloskokov:2005sd}
\bibinfo{author}{S.~Goloskokov}, \bibinfo{author}{P.~Kroll},
  \bibinfo{journal}{Eur.Phys.J.} \bibinfo{volume}{C42} (\bibinfo{year}{2005})
  \bibinfo{pages}{281--301}. \DOIprefix\doi{10.1140/epjc/s2005-02298-5}.
  \href{http://arxiv.org/abs/hep-ph/0501242}{\tt arXiv:hep-ph/0501242}.
%Type = Article
\bibitem[{Polyakov and Semenov-Tian-Shansky(2009)}]{Polyakov:2008aa}
\bibinfo{author}{M.~V. Polyakov}, \bibinfo{author}{K.~M. Semenov-Tian-Shansky},
  \bibinfo{journal}{Eur.Phys.J.} \bibinfo{volume}{A40} (\bibinfo{year}{2009})
  \bibinfo{pages}{181--198}. \DOIprefix\doi{10.1140/epja/i2008-10759-2}.
  \href{http://arxiv.org/abs/0811.2901}{\tt arXiv:0811.2901}.
%Type = Article
\bibitem[{Kumeri\v{c}ki and Mueller(2010)}]{Kumericki:2009uq}
\bibinfo{author}{K.~Kumeri\v{c}ki}, \bibinfo{author}{D.~Mueller},
  \bibinfo{journal}{Nucl.Phys.} \bibinfo{volume}{B841} (\bibinfo{year}{2010})
  \bibinfo{pages}{1--58}. \DOIprefix\doi{10.1016/j.nuclphysb.2010.07.015}.
  \href{http://arxiv.org/abs/0904.0458}{\tt arXiv:0904.0458}.
%Type = Article
\bibitem[{Goldstein et~al.(2011)Goldstein, Hernandez, and
  Liuti}]{Goldstein:2010gu}
\bibinfo{author}{G.~R. Goldstein}, \bibinfo{author}{J.~O.~G. Hernandez},
  \bibinfo{author}{S.~Liuti}, \bibinfo{journal}{Phys.Rev.}
  \bibinfo{volume}{D84} (\bibinfo{year}{2011}) \bibinfo{pages}{034007}.
  \DOIprefix\doi{10.1103/PhysRevD.84.034007}.
  \href{http://arxiv.org/abs/1012.3776}{\tt arXiv:1012.3776}.
%Type = Article
\bibitem[{Mezrag et~al.(2013)Mezrag, Moutarde, and Sabati\'e}]{Mezrag:2013mya}
\bibinfo{author}{C.~Mezrag}, \bibinfo{author}{H.~Moutarde},
  \bibinfo{author}{F.~Sabati\'e}, \bibinfo{journal}{Phys.Rev.}
  \bibinfo{volume}{D88} (\bibinfo{year}{2013}) \bibinfo{pages}{014001}.
  \DOIprefix\doi{10.1103/PhysRevD.88.014001}.
  \href{http://arxiv.org/abs/1304.7645}{\tt arXiv:1304.7645}.
%Type = Article
\bibitem[{Mezrag et~al.(2016)Mezrag, Moutarde, and
  Rodriguez-Quintero}]{Mezrag:2016hnp}
\bibinfo{author}{C.~Mezrag}, \bibinfo{author}{H.~Moutarde},
  \bibinfo{author}{J.~Rodriguez-Quintero}, \bibinfo{journal}{Few Body Syst.}
  \bibinfo{volume}{57} (\bibinfo{year}{2016}) \bibinfo{pages}{729--772}.
  \DOIprefix\doi{10.1007/s00601-016-1119-8}.
  \href{http://arxiv.org/abs/1602.07722}{\tt arXiv:1602.07722}.
%Type = Article
\bibitem[{Fanelli et~al.(2016)Fanelli, Pace, Romanelli, Salme, and
  Salmistraro}]{Fanelli:2016aqc}
\bibinfo{author}{C.~Fanelli}, \bibinfo{author}{E.~Pace},
  \bibinfo{author}{G.~Romanelli}, \bibinfo{author}{G.~Salme},
  \bibinfo{author}{M.~Salmistraro}, \bibinfo{journal}{Eur. Phys. J.}
  \bibinfo{volume}{C76} (\bibinfo{year}{2016}) \bibinfo{pages}{253}.
  \DOIprefix\doi{10.1140/epjc/s10052-016-4101-1}.
  \href{http://arxiv.org/abs/1603.04598}{\tt arXiv:1603.04598}.
%Type = Article
\bibitem[{Dorokhov et~al.(2011)Dorokhov, Broniowski, and
  Ruiz~Arriola}]{Dorokhov:2011ew}
\bibinfo{author}{A.~E. Dorokhov}, \bibinfo{author}{W.~Broniowski},
  \bibinfo{author}{E.~Ruiz~Arriola}, \bibinfo{journal}{Phys.Rev.}
  \bibinfo{volume}{D84} (\bibinfo{year}{2011}) \bibinfo{pages}{074015}.
  \DOIprefix\doi{10.1103/PhysRevD.84.074015}.
  \href{http://arxiv.org/abs/1107.5631}{\tt arXiv:1107.5631}.
%Type = Article
\bibitem[{Broniowski et~al.(2008)Broniowski, Ruiz~Arriola, and
  Golec-Biernat}]{Broniowski:2007si}
\bibinfo{author}{W.~Broniowski}, \bibinfo{author}{E.~Ruiz~Arriola},
  \bibinfo{author}{K.~Golec-Biernat}, \bibinfo{journal}{Phys.Rev.}
  \bibinfo{volume}{D77} (\bibinfo{year}{2008}) \bibinfo{pages}{034023}.
  \DOIprefix\doi{10.1103/PhysRevD.77.034023}.
  \href{http://arxiv.org/abs/0712.1012}{\tt arXiv:0712.1012}.
%Type = Article
\bibitem[{Tiburzi and Verma(2017)}]{Tiburzi:2017brq}
\bibinfo{author}{B.~C. Tiburzi}, \bibinfo{author}{G.~Verma},
  \bibinfo{journal}{Phys. Rev.} \bibinfo{volume}{D96} (\bibinfo{year}{2017})
  \bibinfo{pages}{034020}. \DOIprefix\doi{10.1103/PhysRevD.96.034020}.
  \href{http://arxiv.org/abs/1706.05849}{\tt arXiv:1706.05849}.
%Type = Article
\bibitem[{Hwang and Mueller(2008)}]{Hwang:2007tb}
\bibinfo{author}{D.~Hwang}, \bibinfo{author}{D.~Mueller},
  \bibinfo{journal}{Phys.Lett.} \bibinfo{volume}{B660} (\bibinfo{year}{2008})
  \bibinfo{pages}{350--359}. \DOIprefix\doi{10.1016/j.physletb.2008.01.014}.
  \href{http://arxiv.org/abs/0710.1567}{\tt arXiv:0710.1567}.
%Type = Article
\bibitem[{Chouika et~al.(2017)Chouika, Mezrag, Moutarde, and
  Rodríguez-Quintero}]{Chouika:2017dhe}
\bibinfo{author}{N.~Chouika}, \bibinfo{author}{C.~Mezrag},
  \bibinfo{author}{H.~Moutarde}, \bibinfo{author}{J.~Rodríguez-Quintero},
  \bibinfo{journal}{Eur. Phys. J.} \bibinfo{volume}{C77} (\bibinfo{year}{2017})
  \bibinfo{pages}{906}. \DOIprefix\doi{10.1140/epjc/s10052-017-5465-6}.
  \href{http://arxiv.org/abs/1711.05108}{\tt arXiv:1711.05108}.
%Type = Article
\bibitem[{Müller(2017)}]{Muller:2017wms}
\bibinfo{author}{D.~Müller}  (\bibinfo{year}{2017}).
  \href{http://arxiv.org/abs/1711.09932}{\tt arXiv:1711.09932}.
%Type = Article
\bibitem[{Nakanishi(1969)}]{Nakanishi:1969ph}
\bibinfo{author}{N.~Nakanishi}, \bibinfo{journal}{Prog.Theor.Phys.Suppl.}
  \bibinfo{volume}{43} (\bibinfo{year}{1969}) \bibinfo{pages}{1--81}.
  \DOIprefix\doi{10.1143/PTPS.43.1}.
%Type = Article
\bibitem[{Nakanishi(1963)}]{Nakanishi:1963zz}
\bibinfo{author}{N.~Nakanishi}, \bibinfo{journal}{Phys.Rev.}
  \bibinfo{volume}{130} (\bibinfo{year}{1963}) \bibinfo{pages}{1230--1235}.
  \DOIprefix\doi{10.1103/PhysRev.130.1230}.
%Type = Article
\bibitem[{Mezrag et~al.(2014)Mezrag, Chang, Moutarde, Roberts,
  Rodríguez-Quintero et~al.}]{Mezrag:2014jka}
\bibinfo{author}{C.~Mezrag}, \bibinfo{author}{L.~Chang},
  \bibinfo{author}{H.~Moutarde}, \bibinfo{author}{C.~Roberts},
  \bibinfo{author}{J.~Rodríguez-Quintero}, et~al.,
  \bibinfo{journal}{Phys.Lett.} \bibinfo{volume}{B741} (\bibinfo{year}{2014})
  \bibinfo{pages}{190--196}. \DOIprefix\doi{10.1016/j.physletb.2014.12.027}.
  \href{http://arxiv.org/abs/1411.6634}{\tt arXiv:1411.6634}.
%Type = Article
\bibitem[{Polyakov(1999)}]{Polyakov:1998ze}
\bibinfo{author}{M.~V. Polyakov}, \bibinfo{journal}{Nucl.Phys.}
  \bibinfo{volume}{B555} (\bibinfo{year}{1999}) \bibinfo{pages}{231}.
  \DOIprefix\doi{10.1016/S0550-3213(99)00314-4}.
  \href{http://arxiv.org/abs/hep-ph/9809483}{\tt arXiv:hep-ph/9809483}.
%Type = Article
\bibitem[{Brodsky et~al.(1998)Brodsky, Pauli, and Pinsky}]{Brodsky:1997de}
\bibinfo{author}{S.~J. Brodsky}, \bibinfo{author}{H.-C. Pauli},
  \bibinfo{author}{S.~S. Pinsky}, \bibinfo{journal}{Phys. Rept.}
  \bibinfo{volume}{301} (\bibinfo{year}{1998}) \bibinfo{pages}{299--486}.
  \DOIprefix\doi{10.1016/S0370-1573(97)00089-6}.
  \href{http://arxiv.org/abs/hep-ph/9705477}{\tt arXiv:hep-ph/9705477}.
%Type = Article
\bibitem[{Diehl et~al.(2001)Diehl, Feldmann, Jakob, and Kroll}]{Diehl:2000xz}
\bibinfo{author}{M.~Diehl}, \bibinfo{author}{T.~Feldmann},
  \bibinfo{author}{R.~Jakob}, \bibinfo{author}{P.~Kroll},
  \bibinfo{journal}{Nucl.Phys.} \bibinfo{volume}{B596} (\bibinfo{year}{2001})
  \bibinfo{pages}{33--65}. \DOIprefix\doi{10.1016/S0550-3213(00)00684-2}.
  \href{http://arxiv.org/abs/hep-ph/0009255}{\tt arXiv:hep-ph/0009255}.
%Type = Article
\bibitem[{Pire et~al.(1999)Pire, Soffer, and Teryaev}]{Pire:1998nw}
\bibinfo{author}{B.~Pire}, \bibinfo{author}{J.~Soffer},
  \bibinfo{author}{O.~Teryaev}, \bibinfo{journal}{Eur.Phys.J.}
  \bibinfo{volume}{C8} (\bibinfo{year}{1999}) \bibinfo{pages}{103--106}.
  \DOIprefix\doi{10.1007/s100529901063}.
  \href{http://arxiv.org/abs/hep-ph/9804284}{\tt arXiv:hep-ph/9804284}.
%Type = Article
\bibitem[{Pobylitsa(2002{\natexlab{a}})}]{Pobylitsa:2001nt}
\bibinfo{author}{P.~Pobylitsa}, \bibinfo{journal}{Phys.Rev.}
  \bibinfo{volume}{D65} (\bibinfo{year}{2002}{\natexlab{a}})
  \bibinfo{pages}{077504}. \DOIprefix\doi{10.1103/PhysRevD.65.077504}.
  \href{http://arxiv.org/abs/hep-ph/0112322}{\tt arXiv:hep-ph/0112322}.
%Type = Article
\bibitem[{Pobylitsa(2002{\natexlab{b}})}]{Pobylitsa:2002gw}
\bibinfo{author}{P.~Pobylitsa}, \bibinfo{journal}{Phys.Rev.}
  \bibinfo{volume}{D65} (\bibinfo{year}{2002}{\natexlab{b}})
  \bibinfo{pages}{114015}. \DOIprefix\doi{10.1103/PhysRevD.65.114015}.
  \href{http://arxiv.org/abs/hep-ph/0201030}{\tt arXiv:hep-ph/0201030}.
%Type = Article
\bibitem[{Müller and Hwang(2014)}]{Mueller:2014tqa}
\bibinfo{author}{D.~Müller}, \bibinfo{author}{D.~S. Hwang}
  (\bibinfo{year}{2014}). \href{http://arxiv.org/abs/1407.1655}{\tt
  arXiv:1407.1655}.
%Type = Article
\bibitem[{Hertle(1983)}]{Hertle:1983}
\bibinfo{author}{A.~Hertle}, \bibinfo{journal}{Mathematische Zeitschrift}
  \bibinfo{volume}{184} (\bibinfo{year}{1983}) \bibinfo{pages}{165--192}.
%Type = Article
\bibitem[{Teryaev(2001)}]{Teryaev:2001qm}
\bibinfo{author}{O.~Teryaev}, \bibinfo{journal}{Phys.Lett.}
  \bibinfo{volume}{B510} (\bibinfo{year}{2001}) \bibinfo{pages}{125--132}.
  \DOIprefix\doi{10.1016/S0370-2693(01)00564-0}.
  \href{http://arxiv.org/abs/hep-ph/0102303}{\tt arXiv:hep-ph/0102303}.
%Type = Article
\bibitem[{Radon(1986)}]{Radon:1917tr}
\bibinfo{author}{J.~Radon}, \bibinfo{journal}{Medical Imaging, IEEE
  Transactions on} \bibinfo{volume}{5} (\bibinfo{year}{1986})
  \bibinfo{pages}{170--176}. \DOIprefix\doi{10.1109/TMI.1986.4307775}.
%Type = Article
\bibitem[{Tiburzi(2004)}]{Tiburzi:2004qr}
\bibinfo{author}{B.~Tiburzi}, \bibinfo{journal}{Phys.Rev.}
  \bibinfo{volume}{D70} (\bibinfo{year}{2004}) \bibinfo{pages}{057504}.
  \DOIprefix\doi{10.1103/PhysRevD.70.057504}.
  \href{http://arxiv.org/abs/hep-ph/0405211}{\tt arXiv:hep-ph/0405211}.
%Type = Article
\bibitem[{Polyakov and Weiss(1999)}]{Polyakov:1999gs}
\bibinfo{author}{M.~V. Polyakov}, \bibinfo{author}{C.~Weiss},
  \bibinfo{journal}{Phys.Rev.} \bibinfo{volume}{D60} (\bibinfo{year}{1999})
  \bibinfo{pages}{114017}. \DOIprefix\doi{10.1103/PhysRevD.60.114017}.
  \href{http://arxiv.org/abs/hep-ph/9902451}{\tt arXiv:hep-ph/9902451}.
%Type = Article
\bibitem[{Pobylitsa(2003)}]{Pobylitsa:2002vi}
\bibinfo{author}{P.~Pobylitsa}, \bibinfo{journal}{Phys.Rev.}
  \bibinfo{volume}{D67} (\bibinfo{year}{2003}) \bibinfo{pages}{034009}.
  \DOIprefix\doi{10.1103/PhysRevD.67.034009}.
  \href{http://arxiv.org/abs/hep-ph/0210150}{\tt arXiv:hep-ph/0210150}.
%Type = Article
\bibitem[{Chang et~al.(2013)Chang, Cloet, Cobos-Martinez, Roberts, Schmidt
  et~al.}]{Chang:2013pq}
\bibinfo{author}{L.~Chang}, \bibinfo{author}{I.~Cloet},
  \bibinfo{author}{J.~Cobos-Martinez}, \bibinfo{author}{C.~Roberts},
  \bibinfo{author}{S.~Schmidt}, et~al., \bibinfo{journal}{Phys.Rev.Lett.}
  \bibinfo{volume}{110} (\bibinfo{year}{2013}) \bibinfo{pages}{132001}.
  \DOIprefix\doi{10.1103/PhysRevLett.110.132001}.
  \href{http://arxiv.org/abs/1301.0324}{\tt arXiv:1301.0324}.
%Type = Article
\bibitem[{Yuan(2004)}]{Yuan:2003fs}
\bibinfo{author}{F.~Yuan}, \bibinfo{journal}{Phys.Rev.} \bibinfo{volume}{D69}
  (\bibinfo{year}{2004}) \bibinfo{pages}{051501}.
  \DOIprefix\doi{10.1103/PhysRevD.69.051501}.
  \href{http://arxiv.org/abs/hep-ph/0311288}{\tt arXiv:hep-ph/0311288}.
%Type = Article
\bibitem[{Chang et~al.(2014)Chang, Mezrag, Moutarde, Roberts,
  Rodriguez-Quintero et~al.}]{Chang:2014lva}
\bibinfo{author}{L.~Chang}, \bibinfo{author}{C.~Mezrag},
  \bibinfo{author}{H.~Moutarde}, \bibinfo{author}{C.~D. Roberts},
  \bibinfo{author}{J.~Rodriguez-Quintero}, et~al.,
  \bibinfo{journal}{Phys.Lett.} \bibinfo{volume}{B737} (\bibinfo{year}{2014})
  \bibinfo{pages}{23--29}. \DOIprefix\doi{10.1016/j.physletb.2014.08.009}.
  \href{http://arxiv.org/abs/1406.5450}{\tt arXiv:1406.5450}.
%Type = Article
\bibitem[{Beringer et~al.(2012)}]{Beringer:1900zz}
\bibinfo{author}{J.~Beringer}, et~al. (\bibinfo{collaboration}{Particle Data
  Group}), \bibinfo{journal}{Phys. Rev.} \bibinfo{volume}{D86}
  (\bibinfo{year}{2012}) \bibinfo{pages}{010001}.
  \DOIprefix\doi{10.1103/PhysRevD.86.010001}.
%Type = Article
\bibitem[{Efremov and Radyushkin(1980)}]{Efremov:1978rn}
\bibinfo{author}{A.~Efremov}, \bibinfo{author}{A.~Radyushkin},
  \bibinfo{journal}{Theor.Math.Phys.} \bibinfo{volume}{42}
  (\bibinfo{year}{1980}) \bibinfo{pages}{97--110}.
  \DOIprefix\doi{10.1007/BF01032111}.
%Type = Article
\bibitem[{Lepage and Brodsky(1980)}]{Lepage:1980fj}
\bibinfo{author}{G.~P. Lepage}, \bibinfo{author}{S.~J. Brodsky},
  \bibinfo{journal}{Phys.Rev.} \bibinfo{volume}{D22} (\bibinfo{year}{1980})
  \bibinfo{pages}{2157}. \DOIprefix\doi{10.1103/PhysRevD.22.2157}.
%Type = Article
\bibitem[{Maris et~al.(1998)Maris, Roberts, and Tandy}]{Maris:1997hd}
\bibinfo{author}{P.~Maris}, \bibinfo{author}{C.~D. Roberts},
  \bibinfo{author}{P.~C. Tandy}, \bibinfo{journal}{Phys.Lett.}
  \bibinfo{volume}{B420} (\bibinfo{year}{1998}) \bibinfo{pages}{267--273}.
  \DOIprefix\doi{10.1016/S0370-2693(97)01535-9}.
  \href{http://arxiv.org/abs/nucl-th/9707003}{\tt arXiv:nucl-th/9707003}.
%Type = Article
\bibitem[{Qin et~al.(2014)Qin, Roberts, and Schmidt}]{Qin:2014vya}
\bibinfo{author}{S.-X. Qin}, \bibinfo{author}{C.~D. Roberts},
  \bibinfo{author}{S.~M. Schmidt}, \bibinfo{journal}{Phys. Lett.}
  \bibinfo{volume}{B733} (\bibinfo{year}{2014}) \bibinfo{pages}{202--208}.
  \DOIprefix\doi{10.1016/j.physletb.2014.04.041}.
  \href{http://arxiv.org/abs/1402.1176}{\tt arXiv:1402.1176}.
%Type = Article
\bibitem[{de~Melo et~al.(2006)de~Melo, Frederico, Pace, and
  Salme}]{deMelo:2005cy}
\bibinfo{author}{J.~P. B.~C. de~Melo}, \bibinfo{author}{T.~Frederico},
  \bibinfo{author}{E.~Pace}, \bibinfo{author}{G.~Salme},
  \bibinfo{journal}{Phys. Rev.} \bibinfo{volume}{D73} (\bibinfo{year}{2006})
  \bibinfo{pages}{074013}. \DOIprefix\doi{10.1103/PhysRevD.73.074013}.
  \href{http://arxiv.org/abs/hep-ph/0508001}{\tt arXiv:hep-ph/0508001}.
%Type = Article
\bibitem[{Frederico et~al.(2009)Frederico, Pace, Pasquini, and
  Salme}]{Frederico:2009fk}
\bibinfo{author}{T.~Frederico}, \bibinfo{author}{E.~Pace},
  \bibinfo{author}{B.~Pasquini}, \bibinfo{author}{G.~Salme},
  \bibinfo{journal}{Phys.Rev.} \bibinfo{volume}{D80} (\bibinfo{year}{2009})
  \bibinfo{pages}{054021}. \DOIprefix\doi{10.1103/PhysRevD.80.054021}.
  \href{http://arxiv.org/abs/0907.5566}{\tt arXiv:0907.5566}.
%Type = Article
\bibitem[{Huber et~al.(2008)}]{Huber:2008id}
\bibinfo{author}{G.~Huber}, et~al. (\bibinfo{collaboration}{Jefferson Lab}),
  \bibinfo{journal}{Phys.Rev.} \bibinfo{volume}{C78} (\bibinfo{year}{2008})
  \bibinfo{pages}{045203}. \DOIprefix\doi{10.1103/PhysRevC.78.045203}.
  \href{http://arxiv.org/abs/0809.3052}{\tt arXiv:0809.3052}.

\end{thebibliography}

\end{document}